 \documentclass[12pt]{article}
\def\refnew#1{(\ref{#1})}
\newcommand{\be}{\begin{equation}}
\newcommand{\ee}{\end{equation}}
\newcommand{\ba}{\begin{eqnarray}}
\newcommand{\ea}{\end{eqnarray}}

\newcommand{\inc}{{\it i}}
\newcommand{\efbold}{\mbox{{\boldmath $\vec f$}}}

\newcommand{\erbold}{\mbox{{\boldmath $\vec r$}}}
\newcommand{\Phibold}{\mbox{{\boldmath $\vec \Phi$}}}
\newcommand{\mubold}{\mbox{{\boldmath $\vec \mu$}}}
\newcommand{\etabold}{\mbox{{\boldmath $\vec \eta$}}}

\newcommand{\f}{{\mbox{\boldmath$\vec{f}$}}}
\newcommand{\gbold}{\mbox{{\boldmath $\vec{g}$}}}
\begin{document}
\date{}
\title{ % ~~~~${~}^{~}$\\
 ~~~~~~~~~~~~~~~~~~~~~~~~~~~~~~~~~~~~~~~~~
 ~${~}^{{Submitted~to~\it
{"Astronomy~and~Astrophysics"}}}$
 ~${~}^{~}$\\ {\Large{\bf Gauge Freedom\\ in the
 N-body problem of Celestial Mechanics.}}\\
%{~}^{~}\\
%{~}^{~}\\
}
\author{
 {\Large{Michael Efroimsky}}\\
 {\small{US Naval Observatory, Washington DC 20392 USA}}\\
 {\small{e-mail: efroimsk @ ima.umn.edu~}}\\
~\\
and\\
~\\
 {\Large{Peter Goldreich}}\\
{\small{IAS, ~Princeton NY 08540
 ~\&~ CalTech, ~Pasadena CA 91125 USA}}\\
{\small{e-mail: pmg @ tapir.caltech.edu~}}
 }

 \maketitle
 \begin{abstract}
We summarise research reported in (Efroimsky 2002, 2003; Efroimsky
\& Goldreich 2003a,b) and develop its application to planetary
equations in non-inertial frames.

Whenever a standard system of six planetary equations (in the
Lagrange, Delaunay, or other form) is employed, the trajectory
resides on a 9(N-1)-dimensional submanifold of the
12(N-1)-dimensional space spanned by the orbital elements and
their time derivatives. The freedom in choosing this submanifold
reveals an internal symmetry inherent in the description of the
trajectory by orbital elements. This freedom is analogous to the
gauge invariance of electrodynamics. In traditional derivations of
the planetary equations this freedom is removed by hand through
the introduction of the Lagrange constraint, either explicitly (in
the variation-of-parameters method) or implicitly (in the
Hamilton-Jacobi approach). This constraint imposes the condition
that the orbital elements osculate the trajectory, i.e., that both
the instantaneous position and velocity be fit by a Keplerian
ellipse (or hyperbola). Imposition of any supplementary constraint
different from that of Lagrange (but compatible with the equations
of motion) would alter the mathematical form of the planetary
equations without affecting the physical trajectory.

For coordinate-dependent perturbations, any gauge different from that
of Lagrange makes the Delaunay system non-canonical. In a more general
case of disturbances dependent also upon velocities, there exists a
"generalised Lagrange gauge" wherein the Delaunay system is symplectic
(and the orbital elements are osculating in the phase space). This
gauge reduces to the regular Lagrange gauge for perturbations that
are velocity-independent.

We provide a practical example illustrating how the gauge formalism
considerably simplifies the calculation of satellite motion about an
oblate precessing planet.  \\ \\

% KEY WORDS: ~~~~Orbit integration, Lagrange system, Delaunay system,\\
% {${\left. \; \; \right.}^{\left. \; \right.}\;$}~~
%Hidden symmetries, Gauge invariance.\\
% RUNNING HEAD: ~Hidden symmetry of the Lagrange and Delaunay systems

\end{abstract}

% \pagebreak

\section{Introduction}
\label{sec:Intro}

\subsection{~~Prefatory notes}

On the 6-th of November 1766 young geometer Joseph-Louis Lagrange,
\footnote{The real name of the young man invited in 1766 to the
Prussian court was Giuseppe Lodovico Lagrangia. It was only
several years down the road that he became Joseph-Louis Lagrange.}
invited from Turin at d'Alembert's recommendation by King
Friedrich the Second, succeeded Euler as the Director of
Mathematics at the Berlin Academy. Lagrange held the position for
20 years, and this fruitful period of his life was marked by an
avalanche of excellent results, and by three honourable prizes
received by him from the Acad{\'e}mie des Sciences of Paris. All
three prizes (one of which he shared with Euler) were awarded to
Lagrange for his contributions to celestial mechanics. Among these
contributions was a method initially developed by Lagrange for his
studies of planet-perturbed cometary orbits and only later applied
to planetary motion (Lagrange 1778, 1783, 1788, 1808, 1809, 1810).
The method is based on an elegant mathematical novelty, the
variation of parameters emerging in solutions of differential
equations.  The first known instances of this tool being employed
are presented in a paper on Jupiter's and Saturn's mutual
disturbances, submitted by Euler to a competition held by the
French Academy of Sciences (Euler 1748), and in the treatise on
the Lunar motion, published by Euler in 1753 in St.Petersburg
(Euler 1753). However it was Lagrange who revealed the full power
of the method.

Below we shall demonstrate that the equations for the
instantaneous orbital elements possess a hidden symmetry similar
to the gauge symmetry of electrodynamics. Derivation of the
Lagrange system involves a mathematical operation equivalent to
the choice of a specific gauge. As a result, trajectories get
constrained to some 9-dimensional submanifold in the
12-dimensional space constituted by the orbital elements and their
time derivatives. However, the choice of this submanifold is
essentially ambiguous, and this ambiguity gives birth to an internal
symmetry. The symmetry is absent in the 2-body case, but comes
into being in the N-body setting ($N\,\ge \,3$) where each body
follows an ellipse of varying shape whose time evolution contains
an inherent ambiguity.

For a simple illustration of this point imagine two coplanar
ellipses sharing one focus. Suppose they rotate at different rates
in their common plane. Let a planet be located at one of the
intersection points of these ellipses. The values of the elliptic
elements needed to describe its trajectory would depend upon which
ellipse was chosen to parameterise the orbit. Either
set would be equally legitimate and would faithfully describe the
physical trajectory.  Thus we see that there exists an infinite
number of ways of dividing the actual planet's movement into
motion along its orbit and the simultaneous evolution of the
orbit. Although the physical trajectory is unique, its description
(parametrisation in terms of Kepler's elements) is not. A map
between two different (though physically-equivalent) sets of
orbital elements is a symmetry transformation (a gauge
transformation, in physicists' jargon).

Lagrange never dwelled on that point. However, in his treatment
(based on direct application of the method of variation of
parameters) he passingly introduced a convenient mathematical
condition which removed the said ambiguity. This condition and
possible alternatives to it will be the topics of Sections  1 - 3
of this paper.

In 1834 - 1835 Hamilton put forward his theory of canonical
transformations. Several years later this approach was furthered
by Jacobi who brought Hamilton's technique into astronomy and,
thereby, worked out a new method of deriving the planetary
equations (Jacobi 1866), a method that was soon accepted as
standard. Though the mathematical content of the Hamilton-Jacobi
theory is impeccably correct, its application to astronomy
contains a long overlooked aspect that needs attention. This
aspect is: where is the Lagrange constraint tacitly imposed, and
what happens if we impose a different constraint? This issue will
be addressed in Section 4 of our paper.

\subsection{Osculating Elements vs Orbital Elements}
\label{subsec:elements}

We start in the spirit of Lagrange. Before addressing the N-particle
case, Lagrange referred to the reduced 2-body problem,
\begin{eqnarray}
 \nonumber
{{\bf {\ddot {\vec {r}}}}}\;+\;\frac{\mu}{r^2}\; \frac{{{ \bf \vec
r}}}{r}\;=\;0\;\;\;,
\;\;\;\;\;\;\;\;\;\;\;\;\;\;\;\;\;\;\;\;\;\;\;\;\;\;\;\;\\
\label{1}\\
 \nonumber \erbold\;\equiv\;\erbold_{planet}\,-\;\erbold_{sun}
 \;\;\;,\;\;\;\;\;\;\;\;\;\;\;\;\;\mu\;\equiv\;G(m_{planet}\,+
 \,m_{sun})\;\;\;,
\end{eqnarray}
whose generic solution, a Keplerian ellipse or a hyperbola,
 can be expressed, in some fixed Cartesian frame, as
\begin{equation}
    \erbold \;=\; {\efbold} \left(C_1, ... , C_6, \,t \right)\;\;\;,\;\;\;\;\;\;
 \;\;\;\;{\bf \dot{\erbold}} \;=\;{\bf \vec g}\left(C_1, ... , C_6,
 \,t \right) \;\;\;\;\;\;,\;\;\;\;\;\;
 \label{eq:rrdot}
 \label{2}
\end{equation}
where
\begin{equation}
 {{\bf{\vec{g}}}}\;\equiv\;\left(\frac{\partial { {\efbold}}}{
 \partial
 t}\right)_{C=const} .
 \label{eq:grelf}
 \label{3}
\end{equation}
Since the problem (\ref{1}) is constituted by three second-order
differential equations, its general solution naturally contains
six adjustable constants $\;C_i$. At this point it is irrelevant
which particular set of the adjustable parameters is employed. (It
may be, for example, a set of Lagrange or Delaunay orbital
elements or, alternatively, a set of initial values of the
coordinates and velocities.)

Following Lagrange (1808, 1809, 1810), we employ ${\bf{\efbold}}$
as an ansatz for a solution of the N-particle problem, the
disturbing force acting at a particle being denoted by
$\;\Delta{\bf{\vec F}}\,$:\footnote{Our treatment covers
disturbing forces $\Delta{\bf\vec F}({\bf{\vec r}}, \dot{\bf{\vec
r}}, t)$ that are arbitrary vector-valued functions of position,
velocity, and time.}
\begin{equation}
{\bf \ddot{\vec r}}\;+\;\frac{\mu}{r^2}\;\frac{{\bf\vec r
}}{r}\;=\;{\Delta \bf{\vec F}}\;\;\;.
 \label{4}
 \label{eq:DelF}
\end{equation}
Now the "constants" become time dependent:
\begin{equation}
{\bf \vec r }\;=\;{\bf \efbold} \left(C_1(t), ... , C_6(t), \,t
\right)\,\;\;\;,
 \label{eq:rpert}
 \label{5}
\end{equation}
whence the velocity
\begin{equation}
 \frac{d \erbold}{dt}\;=\;\frac{ \partial
{\bf \efbold}}{\partial t}\;+\; \sum_i \;\frac{\partial {\bf
\efbold}}{\partial C_i}\;\frac{d C_i}{d t}\;= \;{\bf \vec g}\;+\;
\sum_i \;\frac{\partial {\bf \efbold}}{\partial C_i}\;\frac{d C_i}{d
t}\;\;\;\;,
 \label{eq:vpert}
 \label{6}
\end{equation}
acquires a new input, $\;\sum ({\partial{\bf \efbold}}/{\partial
C_i})({dC_i}/{dt})\;$.

Substitution of \refnew{eq:rpert} and \refnew{eq:vpert} into the
perturbed equation of motion (\ref{eq:DelF}) leads to three
independent scalar second-order differential equations which contain
one independent parameter, time, and six functions $\;C_i(
t)\;$. These are to be found from the said three equations, and this
makes the essence of the variation-of-parameters (VOP)
method. However, the latter task cannot be accomplished in a unique
way because the number of variables exceeds, by three, the number of
equations. Thence, though the {\it{~physical~}} trajectory (comprised
by the locus of points in the Cartesian frame and by the values of
velocity at each of these points) is unique, its parametrisation
through the orbital elements is ambiguous. This circumstance was
appreciated by Lagrange who amended the system, by hand, with three
independent conditions,
 \begin{equation}
 \sum_i \;\frac{\partial
{\bf \efbold}}{\partial C_i}\;\frac{d C_i}{d t}\;=\;0 \;\;\;,
 \label{eq:lagcons}
 \label{7}
 \end{equation}
and thus made it solvable. His choice of constraints was
motivated by both physical considerations and mathematical
expediency. Since, physically, the set of functions
$\,\left(C_1(t), ... , C_6(t)\right)\,$ can be interpreted as
parameters of an instantaneous ellipse, in a bound-orbit case, or
an instantaneous hyperbola, in a fly-by situation, Lagrange found
it natural to make the instantaneous orbital elements $\;C_i\;$
osculating. His constraint \refnew{eq:lagcons} fixes the
instantaneous ellipse (or hyperbola) defined by $\,\left(C_1(t),
... , C_6(t)\right)\,$ such that, at each moment of time, it
coincides with the unperturbed (two-body) orbit that the body
would follow if the disturbances were to cease instantaneously.
This way, Lagrange restricted his use of the orbital elements to
elements that osculate in the reference frame wherein ansatz
(\ref{5}) is employed. Lagrange never explored alternative
options; he simply imposed (\ref{7}) and used it to derive his
famous system of equations for orbital elements.

Such a restriction, though physically motivated, is completely
arbitrary from the mathematical point of view. While the
imposition of (\ref{7}) considerably simplifies the subsequent
calculations it in no way influences the shape of the physical
trajectory and the rate of motion along it. A choice of any other
supplementary constraint
 \begin{equation}
 \sum_i \;\frac{\partial {\bf \efbold}}{\partial C_i}\;\frac{d
 C_i}{d t}\;=\; {\bf {\vec \Phi}}(C_{1,...,6}\,,\,t)\;\;\;,\;
 \label{eq:gencons}
 \label{8}
 \end{equation}
$\Phibold\,$ being an arbitrary function of time and parameters
$\;C_i$, would lead to the same physical orbit and the same
velocities.\footnote{In principle, one may endow $\,\Phibold\,$
also with dependence upon the parameters' time derivatives of all
orders. This would yield higher-than-first-order time derivatives
of the $\;C_i\;$ in subsequent developments requiring additional
initial conditions, beyond those on ${\bf\vec r}$ and
${\bf\dot{\vec r}}$, to be specified in order to close the system.
We avoid this unnecessary complication by restricting $\Phibold\,$
to be a function of time and the $\;C_i\;$.} Substitution of the
Lagrange constraint (\ref{eq:lagcons}) by its generalisation
(\ref{eq:gencons}) would not influence the motion of the body but
would alter its mathematical description (i.e., would entail
different solutions for the orbital elements). Such invariance of
a physical picture under a change of parametrisation is called
gauge freedom or gauge symmetry. It parallels a similar phenomenon
well known in electrodynamics and, therefore, has similar
consequences. On the one hand, a good choice of gauge often
simplifies solution of the equations of motion. (In Section 3 we
provide a specific application to motion in an accelerated
coordinate system.) On the other hand, a so-called ``gauge shift''
will occur in the course of numerical orbit computation (Murison
\& Efroimsky 2003).

Derivation of the conventional Lagrange and Delaunay planetary
equations by the VOP method incorporates the Lagrange constraint
(Brouwer \& Clemence 1961). Both systems of equations get altered
under a different gauge choice as we now show. The essence of a
derivation suitable for a general choice of gauge starts with
\refnew{eq:vpert} from which the formula for the acceleration
follows:
\begin{equation}
 \frac{d^2 \erbold }{dt^2}\;= \;\frac{\partial \bf \vec g}{\partial
t}\;+\;\sum_i\;\frac{\partial{\bf \vec g}}{\partial C_i}\;\frac{d
C_i}{d t}\;+\;\frac{d{\bf \vec \Phi}}{dt}\;=\; \frac{\partial^2
{\bf{\efbold}}}{\partial^2 t}\;+\;\sum_i\;\frac{\partial{\bf \vec
g}}{\partial C_i }\;\frac{d C_i}{d t}\;+\;\frac{d{\bf \vec
\Phi}}{dt}\;\;.
 \label{9}
 \label{eq:eqmo}
\end{equation}
Together with the equation of motion \refnew{eq:DelF}, it yields:
\begin{equation}
\frac{\partial^2 {\bf {\efbold}} }{\partial
t^2}\;+\;\frac{\mu}{r^2}\;\frac{\bf{\efbold}}{r} \;+\;\sum_i\;
\frac{\partial {\bf \vec g}}{\partial C_i}\;\frac{d C_i}{d t}\;+\;
\frac{d{\bf \vec \Phi}}{dt}\;=\;\Delta{\bf{\vec
F}}\;\;\;\;,\;\;\;\;\;\;r\;\equiv\;|\erbold|\;=\;|{\bf
{\efbold}}|\;\;\;.
 \label{eq:eqmop}
 \label{10}
\end{equation}
The vector function $\;\bf {\efbold}\;$ was from the beginning
introduced as a Keplerian solution to the two-body problem; hence
it must obey the unperturbed equation (\ref{1})$\,$. On these
grounds the above formula simplifies to:
\begin{equation}
 \sum_i\; \frac{\partial {\bf
\vec g}}{\partial C_i}\;\frac{d C_i}{d t}\;=\;{\Delta{\bf{\vec
F}}}\;-\; \frac{d{\bf{\vec{\Phi}}}}{dt}\;\;\;\;.
 \label{eq:eqmosim}
 \label{11}
\end{equation}
This equation describes the perturbed motion in terms of the
orbital elements. Together with constraint (\ref{eq:gencons}) it
constitutes a well-defined system which may be solved with respect
to $\;dC_i/dt\;$. An easy way to do this is to use the elegant
trick suggested by Lagrange: to multiply the equation of motion by
$\;\partial{\bf{\efbold}}/\partial{C_n}\;$ and to multiply the
constraint by $\;-\,\partial{\bf{\vec g} }/\partial{C_n}\;$. The
former operation results in
\begin{equation}
 \frac{\partial \bf{\efbold}}{\partial
C_n} \,\left( \sum_j \, \frac{\partial \bf \vec g}{\partial
C_j}\,\frac{dC_j}{dt} \right)\;=\;\frac{\partial
\bf{\efbold}}{\partial C_n}\;{\;\Delta{\bf{\vec
F}}}\;-\;\frac{\partial \bf{\efbold}}{\partial C_n}\;\frac{d \bf
\vec \Phi}{dt} \;,
 \label{214}
 \label{12}
\end{equation}
while the latter gives
\begin{equation}
  -\;\frac{\partial \bf \vec g}{\partial C_n}\,
  \left( \sum_j\,\frac{\partial \bf \efbold}{\partial C_j}\,
  \frac{dC_j}{dt} \right)\;=\;-\;\frac{\partial \bf \vec
g}{\partial C_n} \;{\bf \vec \Phi} \;\;\;.
 \label{215}
 \label{13}
\end{equation}
Having summed these two equalities, we arrive to:
\begin{equation}
\sum_j\;[C_n\;C_j]\;\frac{dC_j}{dt}\;=\;
 \frac{\partial \bf {\efbold}}{\partial C_n}\;
  {\Delta \bf{\vec F}}\;-\; \frac{\partial{\bf
{\efbold}}}{\partial C_n} \;\frac{d \bf {\vec \Phi}}{dt}\;-\;
\frac{\partial  \bf {\vec g}}{\partial C_n} \;{\bf {\vec \Phi}}
\;\;\;\;,
 \label{217}
 \label{14}
\end{equation}
$[C_n\;C_j]\;$ standing for the unperturbed (i.e., defined as in
the two-body case) Lagrange brackets:
\begin{equation}
[C_n\;C_j]\;\equiv\;\frac{\partial {{\bf{\efbold}}}}{\partial
C_n}\, \frac{\partial {\bf {{\vec g}}}}{\partial
C_j}\,-\,\frac{\partial {{\bf{\efbold}}}}{\partial C_j}\,
\frac{\partial {\bf {{\vec g}}}}{\partial C_n}\;\;\;\;.
 \label{218}
 \label{15}
 \end{equation}
It was agreed above that $\;\Phibold\;$ is a function of time and
of the "constants" $\;C_i\;$, but not of their time derivatives.
Under this convention, the above equation may be shaped into a
more convenient form:
 \begin{equation}
\sum_j\;\left(\,[C_n\;C_j]\;+\;\frac{\partial \bf\efbold}{\partial
C_n}\;\frac{\partial \bf {\vec \Phi}}{\partial C_j}\;
\,\right)\,\frac{dC_j}{dt}\;=\; \frac{\partial \bf
{\efbold}}{\partial C_n}\; {\Delta \bf{\vec F}}\;-\;
\frac{\partial{\bf {\efbold}}}{\partial C_n} \;\frac{\partial \bf
{\vec \Phi}}{\partial t}\;-\; \frac{\partial \bf {\vec
g}}{\partial C_n} \;{\bf {\vec \Phi}} \;\;\;\;.
 \label{general_F}
 \label{16}
 \end{equation}
This is the most general form of the gauge-invariant perturbation
equations of celestial mechanics. In the Lagrange gauge, when the
$\;\Phibold\;$ terms are absent, we can obtain an immediate solution
for the individual $dC_i/dt$ by exploiting the well known expression
for the Poisson-bracket matrix which is inverse to the
Lagrange-bracket one and is offered in the literature for the two-body
problem. (Be mindful that our brackets (\ref{15}) are defined in the
same manner as in the two-body case; they contain only functions
$\,\bf\efbold\,$ and $\,\bf\vec g\,$ that are defined in the
unperturbed, two-body, setting.) In an arbitrary gauge the presence of
the term proportional to $\;\partial{ \Phibold}/\partial C_j\;$ on the
left-hand side of (\ref{general_F}) complicates the solution for
$dC_i/dt$, but only to the extent of requiring the resolution of a set
of six simultaneous linear algebraic equations.

To draw to a close, we again emphasise that, for fixed
interactions and initial conditions, all possible (i.e.,
compatible and sufficient) choices of gauge conditions expressed
by the vector function $\;\Phibold\;$ lead to a physically
equivalent picture. In other words, the real trajectory is
invariant under reparametrisations permitted by the ambiguity of
the choice of gauge. This invariance has the following meaning.
Suppose the equations of motion for $\;C_{1,...,6}\;$, with some
gauge condition $\,{\bf \vec \Phi}\,$ imposed, render the solution
$\;C_{1,...,6}(t) \;$. The same equations of motion, with another
gauge $\,\tilde{\Phibold}\,$ enforced, furnish the solution
$\;{\tilde{C}}_{i}(t)\;$ that has a different functional form.
Despite this difference, both solutions, $\;C_{i}(t)\;$ and
$\;\tilde{C}_{i}(t) \;$, when substituted back in
(\ref{eq:rpert}), yield the same curve $\;\bf{\vec r}(t)\;$ with
the same velocities $\;\dot{\bf{\vec r}}(t)\;$. In mathematics
this situation is called a fibre bundle, and it gives birth to a
1-to-1 map of $\;C_{i}(t)\;$ onto $\;\tilde{C}_{i}(t)\;$, which is
merely a reparametrisation. In physics this map is called a gauge
transformation. The entire set of these reparametrisations
constitute a group of symmetry known as a gauge group.

Just as in electrodynamics, where the fields $\,\bf \vec E\,$ and
$\,\bf \vec B\,$ stay invariant under gradient transformations of the
4-potential $\,A^{\mu}\,$, so the invariance of the orbit implements
itself through the form-invariance of expression (\ref{eq:rpert})
under the afore mentioned map. The vector $\,\erbold\,$ and its full
time derivative $\;\dot{\erbold}\;$, play the role of the physical
fields $\,\bf \vec E\,$ and $\,\bf \vec B\,$, while the Keplerian
coordinates $\,C_1, ... , C_6\,$ play the role of the 4-potential
$\,A^{\mu}\,$.

A comprehensive discussion of all the above-raised issues can be found
in Efroimsky (2002, 2003). Interconnection between the internal
symmetry and multiple time scales in celestial mechanics is touched
upon in Newman \& Efroimsky (2003).

\section{Planetary equations in an arbitrary gauge}

\subsection{Lagrangian and Hamiltonian Perturbations}

We can proceed further by restricting the class of perturbations
we consider to those in which $\,\Delta \bf \vec F\,$ is derivable
from a perturbed Lagrangian. This restricted class is still broad
enough to encompass most applications of celestial-mechanics
perturbation theory.

Let the unperturbed dynamics be described by Lagrangian
$\;L_o({\bf{\vec r}},\,{\bf \dot{\vec r}},\,t)\;=\;{\bf \dot{\vec
r}}^{\left.\,\right. 2}/2\,-\,U({\bf \vec r}\;,\;t)$, canonical
momentum $\;{\bf{\vec{p}}}\,=\,{\bf \dot{\vec r}}\;$ and
Hamiltonian $\;H_o({\bf{\vec r}},\,{\bf{\vec p}},\,t)\,=\,{\bf
{\vec p}}^{\left.\,\right. 2}/2\,+\,U({\bf \vec r}\;,\;t)\;$.
Disturbed motion will be described by the new, perturbed,
functions:
\begin{eqnarray}
 L({\bf{\vec r}},\,{\bf \dot{\vec r}},\,t)\;=\;L_o\;+
 \;\Delta L\;=\;\frac{{\bf{\dot {\vec r}}}^{\left.\,\right. 2}}{2}\;-\;
  U({\bf \vec r},\;t)\;+
  \;\Delta L ( {\bf \vec r},
 \,{ \bf { \dot { \vec {r}}}} ,\,t) \;\;\;,
 ~~~~~~~~~~~~~~~~~~~~~~~~~~~~~~~~~~~~~~~~
 \label{101}
 \label{17}
\end{eqnarray}
\begin{eqnarray}
 {\bf {\vec p}}\;= \;{\bf{\dot {\vec r}}}\;+\; \frac{\partial \,\Delta
 L}{\partial {\bf{\dot{\vec r}}}} \;\;\;\;,
 ~~~~~~~~~~~~~~~~~~~~~~~~~~~~~~~~~~~~~~~~~~~~~~~~~~~~~~~~~~~
 \label{18}
 \label{102}\\
 \nonumber\\ H({\bf{\vec r}},\,{\bf{\vec p}},\,t)\,=\,{\bf \vec
 p}\,{\bf {\dot {\vec r}}}\,-\,L\,= \,\frac{{\bf \vec p}^{\left.
 \,\right. 2}}{2}\,+ \,U\,+\,\Delta H\;\;,\;\;\;\;\;\;\Delta
 H({\bf{\vec r}},\,{\bf{\vec p}},\,t)\,\equiv\,-\,\Delta
 L\,-\,\frac{1}{2}\,\left(\frac{\partial \,\Delta L}{\partial
 {\bf{\dot {\vec r}}}} \right)^2\;\;\;,\;\;
 \label{19}
 \label{103}
\end{eqnarray}
$\Delta H\;$ being introduced as a variation of the functional form,
i.e., as $\;\Delta H\,\equiv\,H({\bf{\vec r}},\,{\bf{\vec p}},\,t) -
H_o({\bf{\vec r}},\,{\bf{\vec p}},\,t) \;$. The Euler-Lagrange
equations written for the perturbed Lagrangian (\ref{101}) will give:
\begin{equation}
 {\bf{\ddot {\vec r}}}\;=\;-\;\frac{\partial U}{\partial {\bf{\vec r}}} \;+\;
\Delta
 {\bf \vec F}\;\;\;\;,
 \label{20}
 \label{104}
\end{equation}
where the new term
\begin{equation}
 \Delta {\bf \vec F}\;\equiv\;\frac{\partial \,\Delta L}{\partial
 {\bf \vec r}}\;-\;\frac{d}{dt}\,\left(\frac{\partial \,\Delta L}{\partial
{\bf{\dot
 {\vec r}}}}\right)\;\;\;\;
 \label{21}
 \label{105}
\end{equation}
is the disturbing force. Expression (\ref{21}) reveals that whenever
the Lagrangian perturbation is velocity independent, it plays the role
of the disturbing function. Generally, though, the disturbing force is
not equal to the gradient of $\,\Delta L\,$ but has an extra term
generated by the velocity dependence.

Examples in which a velocity dependent $\;\Delta L\;$ has been used in
a celestial mechanics setting include: the treatment of inertial
forces in a coordinate system tied to the spin axis of a precessing
planet (Goldreich 1965); the velocity dependent corrections to
Newton's law of gravity in the relativistic two-body problem (Brumberg
1992).

\subsection{Gauge-invariant planetary equations}

When the expression (\ref{105}) for the most general force
emerging within the Lagrangian formalism is substituted into the
gauge-invariant perturbation equation (\ref{217}), it yields:
\begin{equation}
\sum_j\;[C_n\;C_j]\;\frac{dC_j}{dt}\;=\;\frac{\partial \bf
{\efbold}}{\partial C_n}\; \frac{\partial \, \Delta L}{\partial
\erbold}\;-\; \frac{\partial{\bf {\efbold}}}{\partial C_n}
\;\frac{d}{dt}\left({\bf{\vec \Phi}} \,+\,\frac{\partial\, \Delta
 L}{\partial \bf{\dot{\vec r}}}\right)\;-\; \frac{\partial \bf
{\vec g}}{\partial C_n} \;{\bf {\vec \Phi}} \;\;\;\;.
 \label{219}
 \label{22}
\end{equation}
 Since, for a velocity-dependent disturbance, the chain rule
 \begin{equation}
\frac{\partial \,\Delta L}{\partial C_n} \;=\;\frac{\partial \,
\Delta L }{\partial {{{\erbold}}}} \,\frac{\partial{{\bf
{\efbold}}}}{\partial C_n} \;+\;\frac{\partial \,\Delta
L}{\partial {\bf {\dot {\erbold}}}^{
 \left.~\right.} } \,\frac{\partial \bf\dot{\vec r} }{\partial
 C_n}\;=\;
\frac{\partial \,\Delta L}{\partial {{{\erbold}}}}
\,\frac{\partial{{\bf {\efbold}}}}{\partial C_n}
\;+\;\frac{\partial \Delta L}{\partial {\bf {\dot {\erbold}}}^{
 \left.~\right.} } \,\frac{\partial ({\bf{\vec g}}\,+\,\Phibold) }{\partial
 C_n}
\;\;\;,
 \label{220}
 \label{23}
 \end{equation}
takes place, formula (\ref{219}) may be re-shaped into
 \begin{equation}
\sum_j\;[C_n\;C_j]\;\frac{dC_j}{dt}\;=\;\frac{\partial \,\Delta L
}{\partial C_n} \;-\;\frac{\partial \,\Delta L}{\partial {\bf
{\dot {\erbold}}}^{
 \left.~\right.} } \,\;\frac{\partial \Phibold  }{\partial C_n}
\;-\; \left(\frac{\partial{\bf{\efbold}}}{\partial C_n}
 \;\frac{d}{dt}\;-\; \frac{\partial
\bf {\vec g} }{ \partial C_n}\right) \;\left({\bf {\vec \Phi }
}\;+\; \frac{\partial \,\Delta L}{\partial \bf\dot{\vec r}}
\right)\;\;.\;
 \label{221}
 \label{24}
 \end{equation}
Next we group terms so that the gauge function
 $\;\Phibold\;$ everywhere appears added to
  $\;\partial (\Delta L)/\partial {\bf\dot{\vec r}}
 \;$, and bring the term proportional to $dC_j/dt$
to the left hand side of the equation:
 \begin{eqnarray}
 \nonumber
 \sum_j\;\left(\;[C_n\;C_j]\;+
 \;\frac{\partial {\bf\efbold}}{\partial C_n}\;
  \frac{\partial }{\partial C_j}\;
  \left(\frac{\partial \,\Delta
 L}{\partial {\bf\dot{\vec r}}}\;+\;{\Phibold}
  \right)\;\right)
  \frac{dC_j }{dt }\;\;=
~~~~~~~~~~~~~~~~~~~~~~~~~~~~~~~~~~~~~~~~~\\
 \label{2211}
 \label{25}\\
 \nonumber
\frac{\partial }{\partial C_n}\,\left[\Delta
L\,+\,\frac{1}{2}\,\left(\frac{\partial \,\Delta L}{\partial \bf
\dot{\vec r}} \right)^2 \right]
 \;-\;
\left( \frac{\partial \bf \vec g}{\partial C_n}\;+\;\frac{\partial
\bf \efbold}{\partial C_n}\;\frac{\partial}{\partial
t}\;+\;\frac{\partial \,\Delta L}{\partial \bf\dot{\vec
r}}\;\frac{\partial }{\partial C_n} \right)\left(
  {\bf
{\vec \Phi}}\,+\,\frac{\partial \,\Delta L}{\partial \bf\dot{\vec
r }}
 \right)
 \;\;\;.\;\;
 \end{eqnarray}
These modifications help us to recognise the special
nature of the gauge $\;\Phibold\;=\;-\;\partial (\Delta
L)/\partial {\bf\dot{\vec r}}\;$ which will be the subject of
discussion in the next subsection.

Contrast (\ref{25}) with (\ref{16}): while (\ref{general_F})
expresses the VOP method in the most general form it can have in
terms of the disturbing force $\;\Delta {\bf\vec F}({\bf\vec
r},\,{\bf\dot{\vec r}},\,t)\;$, equation (\ref{2211}) renders the
most general form in the language of a Lagrangian perturbation
$\;\Delta L({\bf\vec r},\,{\bf\dot{\vec r}},\,t)$.

The applicability of so generalised planetary equations in analytical
calculations is complicated by the nontrivial nature of the left-hand
sides of (16) and (25). Nevertheless, the structure of these left-hand
sides leaves room for analytical simplification in particular
situations. One such situation is when the gauge is chosen to be
\begin{eqnarray}
\Phibold\;=\;-\;\frac{\partial\,\Delta L}{\partial {\bf\dot{\vec
r}}}\;+\;\mbox{{\boldmath $\vec{\eta}$}}({\bf\vec{r}},\,t)\;\;\;,
 \label{2222}
 \end{eqnarray}
$\mbox{{\boldmath $\vec{\eta}$}}({\bf\vec{r}},\,t)\;$ being an arbitrary
vector function linear in
$\;{\bf\vec{r}}\,$. (It may be, for example, proportional to
$\;{\bf\vec{r}}\;$ or may be equal, say, to a cross product of
$\;{\bf\vec{r}}\;$ by some time-dependent vector.) Under these
circumstances the left-hand side in (25) reduces to the
Lagrange brackets. The situation becomes especially
simple when $\;\partial \Delta L/\partial \bf\dot{\vec{r}}\;$
happens to be linear in $\;{\bf\vec{r}}\;$, in which case we may put
$\;\etabold(\erbold,\,t)\;=\;\partial \Delta L/\partial
\bf\dot{\vec{r}}\;$ and, thus, employ the trivial Lagrange gauge
$\;\Phibold\,=\,0\;$ instead of the generalised Lagrange gauge.
We shall encounter one such example in section 3.4.

As already stressed above, the Lagrange brackets are gauge-invariant
because functions $\,\bf\efbold\,$ and $\,\bf\vec g\,$ used in are
defined within the unperturbed, two-body, problem (\ref{1} - \ref{3})
that lacks gauge freedom. For this reason, one may exploit, to solve
(\ref{25}), the well-known expression for the inverse of matrix
$\;[C_i\;C_j]\;$. Its elements are simplest (and are either zero or
unity) when one chooses as the "constants" the Delaunay
set of orbital variables (Plummer 1918):
\begin{eqnarray}
 \nonumber
{C}_i\;=\;\left\{\;L\,,\;G\,,\;H\,,\;M_o\,,\;\omega\,,\;\Omega\;\right\}~~~~~
~~~~~~~~~~~~~~~~~~~~~~~~~~ \\
 \label{26}\\
 \nonumber
 L\,\equiv\,\sqrt{\mu\,a}\,\;\;,\;\;\;\;\;
G\,\equiv\,\sqrt{\mu\,a\,\left(1\,-\,e^2\right)}\,\;\;,\;\;\;\;
H\,\equiv\,\sqrt{\mu\,a\,\left(1\,-\,e^2\right)}\;\cos
\inc\,\;\;\;,\;\;\;\;
 \end{eqnarray}
where $\;\mu\;\equiv\;G(m_{\tiny{sun}}\,+\,m_{\tiny{planet}})\;$ and
$\;\left(\,e\,,\;a\,,\;M_o\,,\;\omega\,,\;\Omega\,,\;\inc\,\right)\;$
are the Keplerian elements: $\;e\;$ and $\;a\;$ are the eccentricity
and major semiaxis, $\;M_o\;$ is the mean anomaly at epoch, and the
Euler angles $\;\omega\,,\;\Omega\,,\;\inc\;$ are the the argument of
pericentre, the longitude of the ascending node, and the inclination,
respectively.

The simple forms of the Lagrange and Poisson brackets in Delaunay
elements is the proof of these elements' canonicity in the
unperturbed, two-body, problem: the Delaunay elements give birth
to three canonical pairs $\;(Q_i,\,P_i)\;$ corresponding to a
vanishing Hamiltonian:
$\;(L\,,\;-\,M_o)\,,\;(G\,,\;-\,\omega)\,,\;(H\,,\;-\,\Omega)\;$.
In a perturbed setting, when only a position-dependent disturbing
function $\;R({\bf\vec r},\,t)\;$ is "turned on", it can be
expressed through the Lagrangian and Hamiltonian perturbations in
a simple manner, $\;R({\bf\vec r},\,t)\,=\,\Delta L({\bf\vec
r},\,t)\,=\,-\,\Delta H ({\bf\vec r},\,t)\;$, as can be seen from
the formulae presented in the previous subsection. Under these
circumstances, the Delaunay elements remain canonical, provided
the Lagrange gauge is imposed (Brouwer \& Clemence 1961).  This
long known fact can also be derived from our equation (\ref{25}):
if we put $\;\Phibold\,=\,0\;$ and assume $\;\Delta L\;$
velocity-independent, we arrive to
 \begin{eqnarray}
\sum_j\;[C_n\;C_j]\;\frac{dC_j}{dt}\;=\;\;\frac{\partial\, \Delta
L }{\partial C_n}\,\;\;\;\;\;,\;\;
 \label{simple_case}
 \label{27}
 \ea
 where
 \ba
 \Delta L\;=\;\Delta
L\left({\bf\efbold}(C\,,\,t)\,,\;t\,\right)\;=\;
R\left({\bf\efbold}(C\,,\,t)\,,\;t\right)\;=\;-\;\Delta
H\left({\bf\efbold}(C\,,\,t)\,,\;t\,\right)\;\;\;.
 \label{28}
 \end{eqnarray}
This, in its turn, results in the well known Lagrange system of
planetary equations, provided the parameters $\;C_i\;$ are chosen
as the Kepler elements. In case they are chosen as the Delaunay
elements, then (\ref{27}) leads to the standard Delaunay
equations, i.e., to a symplectic system wherein the pairs
$\;(L\,,\;-\,M_o)\,$,
$\;(G\,,\;-\,\omega)\,,\;(H\,,\;-\,\Omega)\;$ again play the role
of canonical variables, but the Hamiltonian, in distinction to the
unperturbed case, no longer vanishes, instead being equal to
$\;\Delta H\;=\;-\;\Delta L\;$.

In our more general case, the perturbation depends also upon
velocities (and, therefore, $\;\Delta L\;$ is no longer equal to
$\;-\,\Delta H\;$). Beside this, the gauge $\;\Phibold\;$ is set
arbitrary. As demonstrated in Efroimsky (2002, 2003), under these
circumstances the gauge-invariant Delaunay-type system is no
longer symplectic.  However, it turns out that this system regains
the canonical form in one special gauge, one that coincides with
the Lagrange gauge when the perturbation bears no velocity
dependence. The issue is explained at length in our previous
papers (Efroimsky \& Goldreich 2003a,b). Here we offer a brief
synopsis of this study.

\subsection{Generalised Lagrange gauge wherein the Delaunay-type\\
 system is canonical}

Equation (\ref{219}) was cast in the shape of (\ref{2211}) not
only to demonstrate the special nature of the gauge
 \begin{eqnarray}
 \Phibold\;=\;-\;\frac{\partial\,\Delta L}{\partial
 {\bf\dot{\vec r}}}\;\;\;,
 \label{special_gauge}
 \label{29}
 \end{eqnarray}
but also to single out the terms in the square brackets on the
right-hand side of (\ref{2211}): together, these terms give
exactly the Hamiltonian perturbation. Thus we come to the
conclusion that in the special gauge (\ref{29}) our equation
(\ref{2211}) simplifies to
 \begin{eqnarray}
 \sum_j\;[C_n\;C_j]\;\frac{dC_j}{dt}\;=\;-\;\frac{\partial\;\Delta
 H }{\partial C_n}\,\;\;.
 \label{simple_form}
 \label{30}
 \end{eqnarray}
As emphasised in the preceding subsection, the gauge invariance of
definition (\ref{15}) enables us to use the standard
(Lagrange-gauge) expressions for $[C_n\,C_j]^{-1}$ to get the
planetary equations from (\ref{30}).

Comparing (\ref{simple_form}) with (\ref{simple_case}), we see that in
the general case of an arbitrary $\Delta L({\bf\vec r},\,{\bf\dot{\vec
r}},\,t)$ one arrives from (\ref{simple_form}) to the same equations
as from (\ref{simple_case}), except that now they contain $\,-\,\Delta
H$ instead of $\Delta L$. When the orbit is parametrised by the
Delaunay variables, those equations take the form:
\begin{eqnarray}
 \frac{dL}{dt}=
\frac{\partial \,\Delta H}{\partial (\,-\,M_o)} \;\;,\;\;\;\;\;\;
\frac{d(\,-\,M_o)}{dt}\;=\,-\,\frac{\partial \,\Delta H}{\partial
L}\;\;\;\;,\;\;\;\;\;\;\;
 \label{Delaunay.12}
 \label{31}
\end{eqnarray}
\begin{eqnarray}
 \frac{dG}{dt}\;=\;\frac{\partial  \,\Delta H}{\partial
(\,-\,\omega )}\;\;\;\;,\;\;\;\;\;\; \frac{d(\,-\,\omega)
}{dt}\;=\,-\,\frac{\partial  \,\Delta H}{\partial
G}\,\;\;\;\;,\;\;\;\;\;
 \label{Delaunay.34}
 \label{32}
\end{eqnarray}
\begin{eqnarray}
 \frac{d H}{dt}\,=\,\frac{\partial  \,\Delta H}{\partial
(\,-\,\Omega )}\,\;\;\;\;,\;\;\;\;\;\; \frac{d (\,-\,\Omega
)}{dt}\,=\,-\,\frac{\partial
 \,\Delta H}{\partial H}\,\;\;\;\;.\;\;\;\;\;\;\;
 \label{Delaunay.56}
 \label{33}
\end{eqnarray}
which is a symplectic system.\footnote{~In this system $\;H\;$
stands not for the Hamiltonian but for one of the Delaunay
elements.}  For this reason we name this special gauge the
"generalised Lagrange gauge".  In any different gauge
$\;\Phibold\;$ the equations for the Delaunay variables would
contain $\;\Phibold$-dependent terms and would not be symplectic.
(Those gauge-invariant equations, for both Lagrange and Delaunay
elements are presented in Efroimsky (2002, 2003) and Efroimsky \&
Goldreich (2003a,b).). This analysis proves the following
{~\underline {\textbf{THEOREM$\,$}}}: {\textbf{~Though the
gauge-invariant equations for Delaunay elements are, generally,
not canonical, they become canonical in the "generalised Lagrange
gauge".}} That this Theorem is not merely a mathematical
coincidence but has deep reasons beneath it will be shown in
Section 4 where the subject is approached from the Hamilton-Jacobi
viewpoint.

The above Theorem gives one example of the gauge formalism being of
use: an appropriate choice of gauge can considerably simplify the
planetary equations (in this particular case, it makes them
canonical).

According to (\ref{18}), the momentum can be written as
 \begin{equation}
 {\bf\vec{p}}\;=\;{\bf\dot{\vec{r}}}\;+ \;\frac{\partial \Delta
 L}{\partial {\bf{\dot{\vec{r}}}}}\;=\;{\bf{\vec{g}}}\;+\;\Phibold \;+
 \;\frac{\partial \Delta L}{\partial {\bf{\dot{\vec{r}}}}}\;\;\;,
 \label{gp}
 \label{34}
\end{equation}
which, in the generalised Lagrange gauge (\ref{29}), simply
reduces to
 \begin{equation}
 {\bf\vec{p}}\;=\;{\bf{\vec{g}}}\;\;\;.\;
 \label{g=p}
\end{equation}
Vector $\;\bf \vec g\;$ was introduced back in (\ref{2} - \ref{3})
to denote the functional dependence of the unperturbed velocity
upon the time and the parameters $\;C_i\;$. In the unperturbed,
two-body, setting this velocity is equal to the momentum
canonically conjugate to the position $\;\bf \vec r\;$ (this is
obvious from (\ref{18}), for zero $\;\Delta L\;$). This way, in
the unperturbed case equality (\ref{g=p}) is fulfilled trivially.
The fact that it remains valid also under perturbation means that,
in the said gauge, the canonical momentum in the disturbed setting
is the same function of time and "constants" as in the
unperturbed, two-body, case. Thus we have established that the
instantaneous Keplerian ellipses (hyperbolae) defined in gauge
(\ref{29}) osculate the trajectory {\bf in phase space}.

Not surprisingly, the generalised Lagrange gauge (\ref{29})
reduces to $\Phibold=0$ in the simple case of velocity-independent
disturbances.

\section{Gauge Freedom and Freedom of Frame Choice}

\subsection{Osculating ellipses described in different frames of reference.}

The essence of the VOP method in celestial mechanics is the
following. A generic two-body-problem solution expressed by
 \begin{eqnarray}
 \erbold\;\;\;=\;\;\;{{\mbox{
 \boldmath$\vec{f}$}}}\left(C,\,t \right)\;\;\;,\;\;\;\;\;
 \;\;\;\;\;\;\;\;\;\;\;\;\;\;
 \label{35}\\
 \nonumber\\
 \left(\frac{\partial{\mbox{\boldmath$\vec{f}$}} }{\partial t}\right)_{C}=\;\;
 {\bf{\vec{g}}}\left(C,\,t \right)
 \;\;\;,\;\;\;\;\;\;\;\;\;\;\;\;\;\;\;\;\;\;\;\;
 \label{36}\\
 \nonumber\\ \nonumber\\ \left(\frac{\partial {\bf{\vec{g}}}}{\partial
 t}\right)_{C}=\;-\;\frac{\mu}{f^2}\;\frac{{\mbox{
 \boldmath$\vec{f}$}}}{f}\;\;\;\;\;,\;\;\;\;\;\;\;\;\;\;\;\;\;\;\;\;\;\;
 \label{37}
 \end{eqnarray}
is employed as an ansatz to solve the disturbed problem:
 \begin{eqnarray}
 {\bf{\vec{r}}}\,&=&\,{{{\f}}}(C(t),\,t)\;\;,~~~~~~~~~~~~~~~~~~~~~~~~~~~~~~
 \label{38}
 \ea
 \ba
 {\bf{\dot{\vec r}}}\,&=&\,\frac{\partial{{{\f}}}}{\partial
 t}\;+\;\frac{\partial{{\f}}}{\partial C_i}\;\frac{d
 C_i}{dt}\;=\;{\bf{\vec g}}\;+\;{\Phibold}  \;\;\;,~~~~~~~~~~~~~~~~~
 \label{39}
 \ea
 \ba
 \nonumber
 {\bf{\ddot{\vec r}}}\,= \,\frac{\partial {\bf{\vec{g}}}}{\partial
 t}\;+\;\frac{\partial {\bf {\vec g}}}{\partial C_i}\;\frac{d
 C_i}{dt}\;+\;\frac{d \Phibold }{dt}~~~~~~~~~~~~~~~~~~~~~~~~~\\
 \label{40}\\
 \nonumber
 {\left. ~ \right.}^{\left. ~ \right.}~~~~~~~~~~~~~~~~~
=\;-\;\frac{\mu}{f^2}\;\frac{{\mbox{
 \boldmath$\vec{f}$}}}{f}\;+\;\frac{\partial {\bf\vec g}}{\partial C_i}
 \;\frac{d
 C_i}{dt}\;+\;\frac{d \Phibold }{dt}\;\;\;.
 \end{eqnarray}
As evident from (\ref{39}), our choice of a particular gauge is
equivalent to decomposition of the physical motion into a movement
with velocity $\;{\bf{\vec g}}\;$ along the instantaneous ellipse (or
hyperbola, in the fly-by case), and a movement associated with the
ellipse's (or hyperbola's) deformation that goes at the rate
$\;\Phibold\;$. It is then tempting to state that a choice of gauge is
equivalent to a choice of an instantaneous comoving reference frame
wherein to describe the motion. Such an interpretation is, however,
incomplete.  Beside the fact that we decouple the physical velocity in
a certain proportion between $\;{\bf{\vec g}}\;$ and $\;\Phibold\;$,
it also matters {\bf which} physical velocity (i.e., velocity relative
to what frame) is decoupled in this proportion. In other words, our
choice of the gauge does not yet exhaust all freedom: we can still
choose {\textbf{in}} {\textbf{what}} {\textbf{frame}} to write ansatz
(\ref{38}).  We may write it in inertial axes or in some accelerated
system. For example, in the case of a satellite orbiting an
accelerated and precessing planet it is {\bf{convenient}} to write the
ansatz for the planet-related position vector.

The above kinematic formulae (\ref{38}) - (\ref{40}) do not yet
contain information about our choice of the reference system in
which we implement the VOP method. This information shows up at
the next stage, when expression (\ref{40}) is inserted into the
dynamical equation of motion
$\;{\bf{\ddot{\erbold}}}\,=\,-\,(\mu\erbold/r^3)\,+\,{\Delta
\bf\vec F}\;$ to yield
 \begin{equation}
 \frac{\partial {\bf {\vec g}}}{\partial C_i}\;\frac{d
 C_i}{dt}\;+\;\frac{d \Phibold }{dt}\;=\;\Delta
{\bf{\vec{F}}}\;=\;\frac{\partial \,\Delta L}{\partial
 {\bf \vec r}}\;-\;\frac{d}{dt}\,\left(\frac{\partial \,
 \Delta L}{\partial
{\bf{\dot {\vec r}}}}\right)\;\;\;.\;\;
 \label{41}
 \end{equation}
Complete information about the reference system in which we put the
VOP method to work (and, therefore, in which we define the orbital
elements $\,C_i\,$) is contained in the expression for the
perturbation force $\;\Delta {\bf{\vec{F}}}\;$. For example, if the
operation is carried out in an inertial coordinate system, $\;\Delta
{\bf{\vec{F}}}\;$ contains physical forces solely. However, if we wish
to implement the VOP approach in a frame moving with a linear
acceleration $\;\bf\vec a\;$, then $\;\Delta {\bf{\vec{F}}}\;$ also
contains the inertial force $\;-\,\bf\vec a\;$. In case this
coordinate system rotates relative to inertial ones at a rate
$\;\mubold\,$, then $\;\Delta {\bf{\vec{F}}}\;$ also includes the
inertial terms $\;\;-\;2\,{\mubold} \, \times \, {\bf {\dot {\vec
{r}}}}\,-\,{\bf {\dot {\mubold}}}\,\times\,{\bf {\vec
{r}}}\,-\,{\mubold}\times({\mubold}\times{\bf {\vec {r}}})\;$. In
considering the motion of a satellite orbiting an oblate precessing
planet it is most reasonable, though not obligatory, to apply the
method (i.e., to define the time derivative) in axes that precess with
the planet. However, this reasonable choice of coordinate system still
leaves us with the freedom of gauge nomination.

\subsection{Relevant Example}

Gauge freedom of the perturbation equations of celestial mechanics
finds an immediate practical implementation in the description of
test particle motion around an precessing oblate planet
(Goldreich, 1965). It is trivial to extend this to account for
acceleration of the planet's centre of mass.

Our starting point is the equation of motion in the inertial
 frame
\begin{equation}
 {\bf{{\vec r}\,}}''\, =\;
  \frac{\partial U}{\partial \bf{\vec r}}\;\; ,
 \label{11.2}
 \label{42}
 \end{equation}
where U is the total gravitational potential and time derivatives
in the inertial axes are denoted by primes.  Suppose that the
planet's spin axis precesses at angular rate $\;\mubold(t)\;$ and
that the acceleration of its centre of mass is given by
$\;{\vec{\bf a}(t)}$. In a coordinate system attached to the
planet's centre of mass and precessing with it, inertial forces
modify the equation of motion so that it assumes the form:
 \begin{equation}
 {\bf{\ddot{\vec {r}}}} \,=\, \frac{\partial U}{\partial
 {\bf{\vec{r}}}} \,-\, 2{\mubold} \, \times \,
 {\bf{\dot{\vec{r}}}}\,-\,{\bf{\dot{\mubold}}}\,\times\,
 {\bf{\vec{r}}}\,-\,{\mubold}\times({\mubold}\times
 {\bf{\vec{r}}})\;-\;{\vec{\bf a}} \;\;,
 \label{11.3}
 \label{43}
 \end{equation}
 time derivatives in the accelerated frame being denoted by
 dots.

 To implement the VOP approach in terms of the orbital
 elements defined in the accelerated frame, we note that the
 disturbing force on the right-hand side of \refnew{11.3} is
 generated according to \refnew{105} by
 \begin{equation}
 \Delta L\left({\bf {\vec r}},\,{\bf{\dot{\vec r}}},\,t
 \right)\,=\, R\;+\;{\bf{\dot {\vec r} } }{\bf \cdot}
 ({\mubold} \times {\bf{\vec{r}}}) \;+\; \frac{1}{2}\;
 ({\mubold} \times {\bf{\vec{r}}}){\bf \cdot}({\mubold}
 \times {\bf{\vec{r}}})\;-\;{\vec{\bf a}\cdot{\vec{\bf r}}}
 \;\;,
 \label{11.13}
 \label{44}
 \end{equation}
where we denote by $R({\bf {\vec r}}\,,\,t )\,$ the
gravitational-potential perturbation (which the perturbation of
the overall gravitational potential U). Since
 \begin{equation}
\frac{\partial \,\Delta L}{\partial {\bf{\dot{r}}}}\;=\;
{\mubold}\times{\bf\vec{r}}\;\;\;,
 \label{eq:dLddotr}
 \label{45}
 \end{equation}
the corresponding Hamiltonian perturbation reads:
 \begin{eqnarray}
 \nonumber
 \Delta H\;=\;-\;\left[\Delta L\;+\; \frac{1}{2}\left(\frac{\partial
 \,\Delta L}{\partial {\bf{\dot{r}}}} \right)^2 \right]~~~~~~~~~~~~~~~~~~~~~~
~~~~~~~~~\\
   \label{eq:Hpert}
 \label{46}\\
 \nonumber
   =\;-\;\left[\,
 R\;+\;{\bf\vec{p}}\cdot
 ({\mubold}\times{\bf\vec{r}}) \;-\;{\vec{\bf a}\cdot {\vec{\bf
 r}}}\;
 \right]\;=\;-\;\left[\,
 R\;+\;({\bf\vec{r}}\times{\bf\vec{p}})\cdot\mubold\;-\;{\vec{\bf a}\cdot
 {\vec{\bf
 r}}}\;
 \right]\;\;,\\
 \nonumber
 \end{eqnarray}
 with vector $\;{\bf\vec{J}}\;=\;{\bf\vec{r}}\times{\bf\vec{p}}\;$
 being the satellite's orbital angular momentum in the inertial frame.

 According to (\ref{34}) and (\ref{45}), the momentum can be
 written as
 \begin{equation}
 {\bf\vec{p}}\;=\;{\bf{\vec{g}}}\;+\;\Phibold \;+
 \;{\mubold}\times{\efbold}\;\;\;,
 \label{47}
 \end{equation}
whence the Hamiltonian perturbation becomes
 \begin{eqnarray}
 \Delta H\;=\;-\;\left[\;R\;+\;\left(
 {\efbold}\times {\bf{\vec{g}}} \right)\cdot \mubold\;+\;
 \left( \Phibold \;+ \;{\mubold}\times{\efbold} \right) \cdot
 \left( {\mubold}\times{\efbold} \right)
     -\;{\vec{\bf a}\cdot {\efbold}}\;\right]\;\;.
 \label{48}
 \end{eqnarray}

\subsection{Elements defined in an accelerated, rotating frame\\
that osculate in the comoving inertial frame}

In this subsection we recall a calculation carried out by Goldreich
(1965) and Brumberg, Evdokimova \& Kochina (1971) and demonstrate that
it may be interpreted as an example of nontrivial gauge fixing.

Let us implement the VOP method in a frame that is accelerating at
rate $\;\bf\vec a\;$ and rotating at angular rate $\;\mubold\;$
relative to some inertial system S. This means that, in the VOP
equation (\ref{41}), $\;\Delta L\;$ is given by formula (\ref{44}) and
$\;\Delta H\;$ by (\ref{48}).

We now choose to describe the motion in the generalised Lagrange gauge
\refnew{special_gauge}, so the expression $\;\left( \Phibold \;+
\;{\mubold}\times{\bf\vec{r}} \right)\;$ on the right-hand side of
(\ref{48}) vanishes (as follows from (\ref{45})), and the
expression for $\;\Delta H\;$ in terms of ${{\efbold}}$ and
${\vec{\bf{g}}}$ has the form:
 \begin{equation}
\Delta H\;=\;-\;\left[\;R({{\efbold}},\,t)+
\mubold\cdot(\efbold\times {\bf {\vec{g}}}) \;-\;
{\vec{\bf{a}}}\cdot{{\efbold}}\;\right]\;\;\;.
 \label{eq:DelHnonosc}
 \label{49}
 \end{equation}
At the same time, the generic expression for the VOP given by
\refnew{2211} simplifies to \refnew{simple_form}. Insertion of
(\ref{49}) therein leads us to
 \begin{equation}
 [C_r\;C_i]\;\frac{dC_i}{dt}\;=\;\frac{\partial}{\partial C_r}\;
 \left[\;R({{\efbold}},\,t)\;+\;\mubold\cdot(\efbold\times {\bf{\vec g}})\;-\;
 {\vec{\bf a}}\cdot{{\efbold}}\;\right] \;\;\;.
 \label{50}
 \end{equation}
 Interestingly, this equation does not contain $\;\bf \dot{\mubold}
 \;$ even though it is valid for non-uniform precession.
% To understand the physical reason behind this absence of
% $\;\bf\dot\mubold\;$ will become clear after we look into some
% property of the velocity and momentum in this gauge.

As explained in subsection 2.3, in the generalised Lagrange gauge the
vector $\;\bf\vec g\;$ is equal to the canonical momentum $\;{\bf\vec
p}\;=\;{\bf\dot{\vec{r}}}\;+\;{\partial\, \Delta L}/{\partial
{\bf\dot{\vec{r}}}}\;$. In the case when the velocity dependence of
$\;\Delta L\;$ is called into being by inertial forces, the momentum
is, according to (\ref{45}),
\begin{equation}
 {\bf\vec p}\;=\;{\bf\dot{\vec{r}}}\;+\;\frac{\partial\, \Delta
 L}{\partial {\bf\dot{\vec{r}}}}\;
 =\;{\bf\dot{\vec{r}}}\;+\;\mubold\,\times\,{\bf\vec{r}}\;\;,
 \label{51}
 \end{equation}
which is the particle's velocity relative to the inertial frame
comoving with the accelerated, rotating frame.  In this sense we may
say that our elements are defined in the accelerated, rotating frame
but osculate in the comoving inertial one.

In the appendix we provide explicit expression for each of the partial
derivatives of $\;\mubold\cdot{\bf\vec{J}}\;$ that appears in the
planetary equations (\ref{50}).

% The fact that in the generalised Lagrange gauge the vector
% $\;\bf\vec g\;$ coincides with the velocity measured in the
% inertial frame explains the absence of $\;\bf\dot\mubold \;$ in
% the planetary equations. If this derivative were present in
% (\ref{50}), then the parameters $\;C_i\;$ would depend upon it,
% and so would $\;{\bf\vec{g}}(C_i\,,\;t\;$. However, the latter is
% impossible because $\;\bf\vec g\;$ coincides with the
% inertial-frame-related velocity that

\subsection{Elements defined in the accelerated, rotating frame,\\
that osculate in this frame}

Here we not only define the elements in the accelerated, rotating
frame, but we also make them osculate in this system, i.e., we make
them satisfy $\,\Phibold=0$. In this gauge, expression (\ref{48})
takes the following form:
 \begin{equation}
 \Delta H\;=\;-\,\left[R({{\efbold}},\,t)\,+\,
 \mubold\cdot(\efbold\times{\bf{\vec g}})\;+\;
 (\mubold\times\efbold)\cdot(\mubold\times\efbold)\;-\;
 {\vec{\bf a}}\cdot{{\efbold}}\right]\;\;\;.
 \label{eq:DelHosc}
 \label{52}
 \end{equation}
 while equation (\ref{25}), after some algebra,\footnote{~Due to
 (\ref{45}), the second term on the left-hand side in (\ref{25}) is
 proportional to $\;[{\partial (\mubold\times\efbold)}/{\partial
 C_j}]\,\dot{C}_j\;=\;\mubold\times\Phibold\;$ and, therefore,
 vanishes.  The second term on the right-hand side simplifies
 in accordance with
 the simple rule $\;{\bf{\vec{A}}}\,\cdot\,
 ({\bf{\vec{B}}}\times{\bf{\vec{C}}})\,=\,({\bf{\vec{C}}}
 \times{\bf{\vec{A}}})\, \cdot\,{\bf{\vec{B}}}\;$. (See Efroimsky \& (Goldreich 2003b).)}
looks like this:
 \begin{eqnarray}
 \nonumber [C_n\;C_i]\;\frac{dC_i}{dt}\;=\;-\;
 \frac{\partial\,\Delta H}{\partial
 C_n}\,~~~~~~~~~~~~~~~~~~~~~~~~~
 \nonumber \\ &&
 \label{53}\\
 \nonumber +\;\mubold\cdot \left(\frac{\partial{\efbold}}{\partial
 C_n}\times {\bf{\vec g}}\;-\;{\efbold}\times \frac{\partial{\vec{\bf
 g}}}{\partial C_n}\right)\;-\; {\bf{\dot{\mubold
 }}}\cdot\left(\efbold\times \frac{\partial \efbold }{\partial
 C_n}\right)\;-\;\left(\mubold\times\efbold\right) \;\frac{\partial
 }{\partial C_n}\left(\mubold\times\efbold\right) \;\;\;.
 \end{eqnarray}
 When substituting (\ref{52}) into (\ref{53}), it is convenient to
rent the expression for $\;\Delta H\;$ apart and to group the term
$\;(\mubold\times\f)\cdot(\mubold\times\efbold)\;$ with the last
term on the right-hand side of (\ref{53}):
 \begin{eqnarray}
 \nonumber [C_n\;C_i]\;\frac{dC_i}{dt}\;=\;
 \frac{\partial}{\partial C_n}\left[R({{\efbold }},\,t)+
 \mubold\cdot(\efbold \times {\bf{\vec g}})
 \;-\; {\vec{\bf a}}\cdot{{\efbold}}\right]~~~~~~~~~~~~~~~~
 \nonumber \\ &&
 \label{eq:avoposc}
 \label{54}\\
  \nonumber
 +\;\mubold\cdot
 \left(\frac{\partial{\efbold}}{\partial C_n}\times
 {\bf{\vec g}}\;-\;{\efbold}\times
 \frac{\partial{\vec{\bf g}}}{\partial C_n}\right)\;-\;
 {\bf{\dot{\mubold }}}\cdot\left(\efbold\times
 \frac{\partial \efbold }{\partial C_n}\right)\;+\;\left(\mubold\times
\efbold\right)
 \;\frac{\partial }{\partial C_n}\left(\mubold\times\efbold\right) \;\;\;.
 \end{eqnarray}
In so writing \refnew{eq:avoposc} we have deliberately cast it
into a form that eases comparison with \refnew{50}.

In the appendix we set up an apparatus from which the partial derivatives
of the inertial terms with respect to the orbital elements may be obtained.
We also show that some of these derivatives vanish. However, a complete
evaluation of the inertial input to the planetary equations in the
ordinary Lagrange gauge involves a long and tedious calculation which we
do not carry out.

\subsection{Comparison of the two gauges}

One of the powers of gauge freedom lies in the availability of
gauge choices that simplify the planetary equations, as we can see
from contrasting (\ref{50}) with (\ref{54}). While the latter
equation is written under the customary Lagrange constraint
(i.e., for elements osculating in the frame where they are
defined), the former equation is written under a nontrivial
constraint called the "generalised Lagrange gauge." The simplicity of
(\ref{50}) speaks for itself.

By identifying the parameters $\;C_i\;$ with the Delaunay
variables, one arrives from (\ref{50}) and (\ref{54}) to the
appropriate Delaunay-type equations (see Appendix I to Efroimsky
\& Goldreich 2003a). The Lagrange equations corresponding to
(\ref{50}) and to (\ref{54}) may be derived from each of these two
equations by choosing $\;C_i\;$ as the Kepler elements and using
the appropriate Lagrange brackets. These Lagrange equations are
written down in the Appendix to Efroimsky \& Goldreich (2003b)
to which we refer the interested reader.

Although the planetary equations are much simpler in the
generalised Lagrange gauge than in the ordinary Lagrange gauge,
some of these differences are less important than others. In many
physical situations, though not always, the $\;{{\mubold\,}^2}\,$
and $\;{\bf{\dot{\mubold}}}\;$ terms in (\ref{54}) are of a higher
order of smallness compared to those linear in $\;\mubold\,$, and
therefore may be neglected, at least for sufficiently short
times.\footnote{~As an example of an exception to this rule, we
mention Venus whose wobble is considerable. This means that, for
example, the $\;{\bf{\dot{\mubold}}}\;$ term cannot be neglected
in computations of circumvenusian orbits.}

\section{Planetary Equations and Gauges in the
         Hamilton-Jacobi Approach}

In this section we demonstrate that the derivation of planetary
equations in the N-particle ($N \,\geq\,3$) case, performed
through the medium of Hamilton-Jacobi method, implicitly contains
a gauge-fixing condition not visible to the naked eye. We present
a squeezed account of our study; a comprehensive description
containing technical details may be found in Efroimsky \&
Goldreich (2003a).

The Hamilton-Jacobi analysis rests on the availability of
different canonical descriptions of the same physical process. Any
two such descriptions, $\;(\,q,\,p,\,H(q,p)\,)\;$ and
$\;(\,Q,\,P,\,H^{*}(Q,P)\,)\;$, correspond to different
parametrisations of the same phase flow, and both obey
Hamilton's equations. Due to the latter circumstance the
infinitesimally small variations
 \begin{equation}
d \theta \,=\,p\,dq\,-\,H\,dt
 \label{425}
 \label{55}
 \end{equation}
and
 \begin{equation}
d \tilde{\theta} \,=\,P\,dQ\,-\,H^{*}\,dt
 \label{426}
 \label{56}
 \end{equation}
are perfect differentials, and so is their difference
 \begin{equation}
 -\,dW\,\equiv\,d \tilde{\theta}\,-\,d
 \theta\;=\;P\,dQ\,-\,p\,dq\,-\,\left(H^{*}\,-\,H\right)\,dt\;\; .
 \label{427}
 \label{57}
 \end{equation}
Here, vectors $\;q,\,p,\,Q,\,$ and $\,P\,$ each contain $N$
components.  Given a phase flow parametrised by a set
$\;(\,q,\,p,\,H(q,p,t)\,)\;$, it is always useful to simplify the
description by a canonical transformation to a new set
$\;(\,Q,\,P,\,H^{*}(Q,P,t)\,)\;$, with the new Hamiltonian $\;H^{*}\;$
being constant in time. Most advantageous are transformations that
nullify the new Hamiltonian $\;H^{*}\;$, because then the new
canonical equations render the variables $\;(\,Q,\,P\,)\;$
constant. A powerful method of generating such transformations stems
from (\ref{427}) being a perfect differential. It is sufficient to
consider $\;W\;$ to be a function of the time and only two other
canonical variables, for example $\;q\;$ and $\;Q\;$. Then (\ref{427})
may be written as
 \begin{equation}
-\,\frac{\partial W}{\partial t}\;dt\;-\;\frac{\partial
W}{\partial Q}\;dQ\;-\;\frac{\partial W}{\partial q}\;dq\;=\;
P\,dQ\;-\;p\,dq\;+\;\left(H\,-\,H^{*}\right)\,dt
 \label{428}
 \label{58}
 \end{equation}
from which it follows that
 \begin{equation}
 P\;=\;-\;\frac{\partial W}{\partial
 Q}\;\;\;\;,\;\;\;\;\;\;\;p\;=\;\frac{\partial W}{\partial
 q}\;\;\;\;,\;\;\;\;\;\;\;H(q,\,p,\,t)\;+\;\frac{\partial W}{\partial
 t}\;=\;H^{*}(Q,\,P,\,t)\;\;\;.
 \label{429}
 \label{59}
 \end{equation}
Inserting the second equation into the third and assuming that
$\;H^{*}(Q,\,P,\,t)\;$ is simply a constant, we get the famous
Jacobi equation
 \begin{equation}
H\left(q,\,\frac{\partial W}{\partial
q}\,,\,t\right)\;+\;\frac{\partial W}{\partial t}\; =\;H^{*}
 \label{430}
 \label{60}
 \end{equation}
whose solution furnishes the transformation-generating function W.
The elegant power of the method becomes most visible if the
constant $\;H^{*}\;$ is set to zero. Under this assumption the
reduced two-body problem is easily resolved. Starting with the
three spherical coordinates and their canonical momenta as
$\;(q,\,p)\;$, one arrives to canonically conjugate constants
$\;(Q,\,P)\;$ that coincide with the Delaunay elements (\ref{26}):
$\;\;(Q_1\,,\;P_1)\,=\,(L\,,\;-\,M_o)\; ;
 \;\;(Q_2\,,\;P_2)\,=\,(G\,,\;-\,\omega)\; ;
 \;\;(Q_3\,,\;P_3)\,=\,(H\,,\;-\,\Omega)\;$.

Extension of this approach to the N-particle problem begins with
consideration of a disturbed 2-body setting. The number of degrees of
freedom is still the same (three coordinates $\,q\,$ and three
conjugate momenta $\,p\,$), but the initial Hamiltonian is perturbed:
\begin{equation}
\dot q\;=\;\frac{\partial (H\,+\,\Delta H)}{\partial
p}\;\;\;,\;\;\;\;\;\;\dot p\;=\;-\;\frac{\partial (H\,+\,\Delta
H)}{\partial q}\;\;.
 \label{449}
 \end{equation}
While in (\ref{428}) - (\ref{430}) one begins with the initial
Hamiltonian $\;H\;$ and ends up with $\;H^{*}\;=\;0\;$, the method may
be extended to the perturbed setting by accepting that now we start
with a disturbed initial Hamiltonian $\;H\,+\,\Delta H\;$ and arrive,
through the same canonical transformation, to an equally disturbed
eventual Hamiltonian $\;H^{*}\,+\,\Delta H\;=\;\Delta H\;$. Plugging
these new Hamiltonians into (\ref{428}) leads to cancellation of
the disturbance $\;\Delta H\;$ on the right-hand side, whereafter one
arrives to the same equation for $\;W(q,\,Q,\,t)\;$ as in the
unperturbed case. Now, however, the new canonical variables are no
longer conserved but obey the dynamical equations
 \begin{equation}
 \dot Q\;=\;\frac{\partial \,\Delta H}{\partial
 P}\;\;\;,\;\;\;\;\;\;\dot P\;=\;-\;\frac{\partial \,\Delta
 H}{\partial Q}\;\;\;\;.
 \label{A4}
 \end{equation}
Because the same generating function is used in the perturbed and
unperturbed cases, the new, perturbed, solution $\;(q,\,p)\;$ is
expressed through the perturbed "constants" $\;Q(t)\,$ and
$\,P(t)\;$ in the same manner as the old, undisturbed, $\;q\;$ and
$\;p\;$ were expressed through the old constants $\;Q\;$ and
$\;P\;$. This form-invariance provides the key to the N-particle
problem: one should choose the transformation-generating function
$\;W\;$ to be additive over the particles and repeat this
procedure for each of the bodies, separately.

Armed with this preparation, we can proceed to uncover the implicit
gauge choice made in using the Hamilton-Jacobi method to derive
evolution equations for the orbital elements. To do this we
substitute the equalities
 \begin{equation}
\dot{ Q}\;=\;\frac{\partial \,\Delta H }{ \partial P }\;=\;\frac{
\partial \,\Delta H}{\partial q }\;\frac{\partial q }{ \partial P }\;
+\;\frac{ \partial \,\Delta H}{\partial p }\;\frac{\partial p }{
\partial P }\;\;
 \label{A5}
 \end{equation}
and
 \begin{equation}
\dot{P }\;=\;-\;\frac{\partial \,\Delta H}{\partial
Q}\;=\;-\;\frac{\partial  \,\Delta H }{ \partial q
}\;\frac{\partial q }{
\partial Q  }\;-\;\frac{\partial  \,\Delta H }{ \partial p  }
\;\frac{\partial p  }{ \partial P  }\;\;
 \label{A6}
 \end{equation}
into the expression for the velocity
 \begin{equation}
\dot{ q }\;=\;\frac{\partial q}{\partial t}\;+\;\frac{\partial q
}{\partial Q }\;\dot{Q}\;+\;\frac{\partial q }{\partial P
}\;\dot{P}\;\; .
 \label{A7}
 \end{equation}
This leads to
 \begin{eqnarray}
 \nonumber
\dot{ q}\;&=&\;\frac{\partial q}{\partial t}\;+\;\left(
\frac{\partial  q}{\partial  Q  }\;\frac{\partial  q}{\partial P
 }\;-\;\frac{\partial q }{\partial  P  }\;\frac{\partial q
 }{\partial Q }\right)\;\frac{\partial \,\Delta H}{\partial q}\;+\;
\left(\frac{\partial  q}{\partial  Q  }\;\frac{\partial  p}{\partial P
 }\;-\;\frac{\partial q }{\partial  P  }\;\frac{\partial p
 }{\partial Q }\right)\;\frac{\partial \,\Delta H}{\partial p} \\
 \label{A8}\\
 \nonumber &=&\;g\;+\;\left( \frac{ \partial \, \Delta H }{ \partial p}
\right)_{q,\,t}\;\;\;,\;\;\;\;\;\;\;\;\;\;\;\;\;g\;\equiv\;
\frac{\partial q}{\partial t}~~~,
\end{eqnarray}
 where we have taken into account that the Jacobian of the canonical
transformation is unity:
\begin{equation}
 \frac{\partial q}{\partial Q}\;\frac{\partial p}{\partial P }\;
-\;\frac{\partial q }{\partial P}\;\frac{\partial p }{\partial Q }\;
=\;1\;\;\;.  \label{A9}
 \end{equation}
To establish the link between the regular VOP method and the
canonical treatment, compare (\ref{A8}) with (\ref{39}). We see
that the symplectic description necessarily imposes a particular
gauge $\;\Phi\;=\;{\partial \,\Delta H}/{\partial p}\;$.

It can be easily demonstrated that this special gauge  coincides
with the generalised Lagrange gauge (\ref{special_gauge})
discussed in subsection 2.2. To that end one has to compare the
Hamilton equation for the perturbed Hamiltonian (\ref{103}),
 \begin{equation}
 \dot{q}\;=\;\frac{\partial \,\left(H\,+\,\Delta H\right)}{\partial
 p}\;=\;p\;+\;\frac{\partial \,\Delta H}{\partial
 p}\;\;\;,
 \label{5558}
 \end{equation}
with the definition of momentum from the Lagrangian (\ref{101}),
 \begin{equation}
 p\,\equiv\,\frac{\partial\,\left(\,
L(q,\,\dot{q},\,t)\,+\,\Delta
L(q,\,\dot{q},\,t)\,\right)}{\partial \dot
q}\;=\;\dot{q}\;+\;\frac{\partial \,\Delta L}{\partial
\dot{q}}\;\;.
 \label{5559}
 \end{equation}
Equating the above two expressions immediately yields
 \begin{equation}
 \Phi\;\equiv\; \left( \frac{ \partial \, \Delta H }{ \partial
p}\right)_{q,\,t}\;=\;-\;\left(\frac{\partial\;\Delta L}{\partial
\dot{q}}\right)_{q,\,t}~~~~~~~~~~~~~~~~~~~~~~~~
 \label{5557}
 \end{equation}
which coincides with (\ref{special_gauge}). Thus the
transformation generated by $\;W(q,\,Q,\,t)\;$ is canonical only
if the physical velocity $\;\dot q\;$ is split in a fashion
prescribed by (\ref{A8}), i.e., if (\ref{5557}) is fulfilled. This
is exactly what our Theorem from subsection 2.2 says.

To summarise, the generalised Lagrange constraint,
$\;\Phibold\,=\,-\;{\partial \,\Delta L}/{\partial \dot q}\;$, is
tacitly instilled into the Hamilton-Jacobi method. Simply by employing
this method (at least, in its straightforward form), we automatically
fix the gauge.\footnote{An explanation of this phenomenon from a
different viewpoint is offered in Section 6 of Efroimsky (2003) where
the Delaunay equations are derived also through a direct change of
variables. It turns out that the outcome retains the symplectic form
only if an extra constraint is imposed by hand.}  By sticking to the
Hamiltonian description we sacrifice gauge freedom.

Above, in subsection 2.3, we established that in the generalised
Lagrange gauge the momentum coincides with $\;\bf\vec g\;$. We now can
get to the same conclusion from (\ref{A8}), (\ref{5559}) and
(\ref{5557}):
 \begin{equation}
 p\,\equiv\,\frac{\partial\,\left(\,
L(q,\,\dot{q},\,t)\,+\,\Delta
L(q,\,\dot{q},\,t)\,\right)}{\partial \dot
q}\;=\;\dot{q}\;-\;\Phi\;=\;g\;\;.
 \label{5555}
 \end{equation}
 Thus, implementation of the Hamilton-Jacobi theory in celestial
 mechanics demands the orbital elements to osculate in phase
 space.  Naturally, this demand reduces to that of regular osculation
 in the simple case of velocity-independent $\;\Delta L \;$.

\section{Conclusions}

In the article thus far we have studied the topic recently raised
in the literature: the planetary equations' internal symmetry that
stems from the freedom of supplementary condition's choice. The
necessity of making such a choice constrains the trajectory
to a 9-dimensional submanifold of the 12-dimensional space
spanned by the orbital elements and their time derivatives.
Similarly to the field theory, the choice of the constraint (= the
choice of gauge) is vastly ambiguous and reveals a hidden symmetry
instilled in the description of the N-body problem in the language
of orbital elements.

We addressed the issue of internal freedom in a sufficiently
general setting where a perturbation to the two-body problem depends
not only upon positions but also upon velocities. Such situations
emerge when relativistic corrections to Newton's law are taken
into account or when the VOP method is employed in rotating
systems of reference.

Just as a choice of an appropriate gauge simplifies solution of
the equations of motion in electrodynamics, an alternative (to
that of Lagrange) choice of gauge in the celestial mechanics can
simplify orbit calculations. We provided one such example,a
satellite orbiting a precessing planet. In this example, the
choice of the generalised Lagrange gauge considerably simplifies
matters.

We have explained where the Lagrange constraint tacitly enters the
Hamilton-Jacobi derivation of the Delaunay equations. This
constraint turns out to be an inseparable (though not easily
visible) part of the method: in the case of momentum-independent
disturbances, the N-body generalisation of the 2-body
Hamilton-Jacobi technique is legitimate only if we use orbital
elements that are osculating. In the situation
where the disturbance depends not only upon positions but also
upon velocities, another constraint (which we call the "generalised
Lagrange constraint") turns out to be stiffly embedded into the
Hamilton-Jacobi development of the problem.

Unless a specific constraint (gauge) is imposed by hand, the
planetary equations assume their general, gauge-invariant, form.
In the case of a velocity-independent disturbance, any gauge
different from that of Lagrange drives the Delaunay system
away from its symplectic form. If we permit the disturbing force
to depend also upon velocities, the Delaunay equations retain
their canonicity only in the generalised Lagrange gauge.
Interestingly, in this special gauge the instantaneous ellipses
(hyperbolae) osculate in phase space.

Briefly speaking, N-body celestial mechanics, expressed in terms of
orbital elements, is a gauge theory but it is not strictly
canonical. It becomes canonical in the generalised Lagrange gauge.\\

~\\
{\bf Acknowledgements}\\
~\\
ME would like to thank Sergei Klioner
% and William Newman
for advice and criticisms. Research by ME was supported by NASA
grant W-19948. Research by PG was partially supported by NSF grant
AST 00-98301.
\\

~\\
{\underline{\bf{\Large{Appendix}}}}\\
~\\

In this appendix we set up an apparatus from which one may
evaluate the partial derivatives with respect to the orbital
elements of inertial terms that appear in the planetary equations
derived in sections 3.3 and 3.4.  We then show that some of these
derivatives vanish. Following that, we derive explicit expressions
for each derivative of $\;\mubold\cdot\left(\efbold\times{\bf\vec
g}\right)\;$, which provides a complete analytic evaluation of the
rotational input in the generalised Lagrange gauge. The topic is
further developed (and the appropriate generalised Lagrange system
of equations is presented) in (Efroimsky \& Goldreich 2003b).

To find the explicit form of the dependence
$\;\efbold\,=\,\efbold(C_i,\,t)\;$, it is conventional to
introduce an auxiliary set of Cartesian coordinates $\;\bf{\vec
q}\;$, with an origin at the gravitating centre, and with the
first two axes located in the plane of orbit. The $\,\bf\vec q\,$
coordinates are easy to express through the major semiaxis
$\,a\,$, the eccentricity $\;e\;$ and the eccentric anomaly
$\;E\;$:
 \ba
q_1\;\equiv\;a\left(\cos
E\;-\;e\right)\;\;,\;\;\;\;\;q_2\;\equiv\;a\sqrt{1\,-\,e^2}\;\sin
E\;\;,\;\;\;\;\;q_3\;=\;0\;\;\; ,
 \label{AA1}
 \ea where $E\;$ itself is a function of the major semiaxis $\;a\,$,
the eccentricity $\;e\,$, the mean anomaly at epoch, $\,M_0\,$,
and the time, $\,t\,$. The time dependence is realised through the
Kepler equation
 \begin{equation}
 E\,-\,e\;\sin E\;=\;M\;\;,
 \label{AA2}
 \end{equation}
where
\begin{equation}
\;M\,\equiv\,M_o\,+\,\mu^{1/2}\,\int_{t_o}^t \;a^{-3/2}\;dt\;\;.
\label{AB1}
\end{equation}
The inertial-frame-related position of the body reads:
 \begin{equation}
{{\erbold}} \;=\;\efbold
\left(\Omega,\,\inc,\,\omega,\,a,\,e,\,M_o\;;\,\,t\right)\;=\;{\bf
\hat{R}}(\Omega,\,\inc,\,\omega)\;\,{\bf {\vec
q}}\left(a,\,e,\;E(a,\,e,\,M_o,\,t)\,\right)\;\;\;,
 \label{AA3}
 \end{equation}
 ${\bf \hat R}(\Omega,\,\inc,\,\omega)\;$ being the matrix of rotation
from the orbital-plane-related coordinate system $\;\bf q\;$ to
the fiducial frame $\;(x,\,y,\,z)\;$ in which the vector $\;\bf
\vec r\;$ is defined. This rotation is parametrised by the three
Euler angles: inclination, $\;\inc\;$; the longitude of the node,
$\;\Omega\;$; and the argument of the pericentre, $\;\omega\,$.

In the unperturbed two-body setting the velocity is expressed by
 \begin{equation}
 {\bf{\vec{g}}}\;=\;\frac{\partial }{\partial t} \;\efbold
 \left(\Omega,\,\inc,\,\omega,\,a,\,e,\,M_o\;;\,\,t\right)\;
 =\;\left(\frac{\partial E}{\partial t}\right)_{a,\,e,\,M_o}{\bf
 \hat{R}}(\Omega,\,\inc,\,\omega)\;\;\left(\frac{\partial {\bf\vec
 q}}{\partial E}\right)_{a,\, e}\;\,\;\;.
 \label{AA4}
 \end{equation}
One can similarly calculate partial derivatives of $\;\efbold\;$
with respect to $\;M_o\;$:
  \begin{equation}
 \frac{\partial }{\partial M_o} \;\efbold
 \left(\Omega,\,\inc,\,\omega,\,a,\,e,\,M_o\;;\,\,t\right)\;
 =\;\left(\frac{\partial E}{\partial M_o}\right)_{a,\,e,\,t}{\bf
 \hat{R}}(\Omega,\,\inc,\,\omega)\;\;\left(\frac{\partial {\bf\vec
 q}}{\partial E}\right)_{a,\, e}\;\,\;\;,
 \label{AA6}
 \end{equation}
whence it becomes evident that $\;\partial \efbold /\partial M_o\;$
is parallel to $\;\bf\vec g\;$ and, hence,
\begin{equation}
{\bf{\vec g}}\;\times\; \left(\frac{\partial \efbold}{\partial
M_o}\right)_{\Omega,\,\inc,\,\omega,\,a,\,e,\,t}\;=\;0\;\;\;.
 \label{AA7}
 \end{equation}
By a similar trick it is possible to demonstrate that
$\;\partial (\efbold\times{\bf\vec g})/\partial M_o\;$ is
proportional to  $\;\partial (\efbold\times{\bf\vec g})/\partial
E\;$ and, therefore, to $\;\partial (\efbold\times{\bf\vec
g})/\partial t\;$. Hence, this derivative vanishes (because in
the two-particle case the cross product $\;\efbold\times{\bf\vec
g}\;$ is an integral of motion). This vanishing of $\;\partial
(\efbold\times{\bf\vec g})/\partial M_o\;$, along with (\ref{AA7}),
implies:
 \begin{equation}
{\bf{\efbold}}\;\times\; \left(\frac{\partial \bf\vec g}{\partial
M_o}\right)_{\Omega,\,\inc,\,\omega,\,a,\,e,\,t}\;=\;0\;\;\;.
 \label{AA8}
 \end{equation}
In the situation when the parameters $\;C_i\;$ are implemented by
the Delaunay elements, a similar sequence of calculations leads to
 \begin{equation}
 {\bf{\vec g}}\;\times\; \left(\frac{\partial \efbold}{\partial
M_o}\right)_{\Omega,\,\omega,\,L,\,G,\,H,\,t}\;=\;0\;\;\;
 \label{AA9}
 \end{equation}
and, appropriately, to
 \begin{equation}
 {\bf{\efbold}}\;\times\; \left(\frac{\partial \bf\vec g}{\partial
M_o}\right)_{\Omega,\,\omega,\,L,\,G,\,H,\,t}\;=\;0\;\;.
 \label{A10}
 \end{equation}

We can proceed much farther in the generalised Lagrange gauge, at
least in so far as derivatives of the rotational input $\;{\bf\vec
J}/\mu\;=\;\efbold\times\gbold\;$ are concerned. (We remind that
here and everywhere $\;\mu\;$ stands for the reduced mass, while
$\;\mubold\;$ denotes the precession rate.)

As we proved above, this cross product is independent of
$\;M_o\;$ and, hence,
\begin{equation}
\mubold\;\cdot\;\frac{\partial(\efbold\times\gbold)}{\partial
M_o}\;=0\;\; .
\end{equation}

Since $\;{\bf\vec J}\;$ is orthogonal to the orbit plane, it is
invariant under rotations of the orbit within its plane, whence
\begin{equation}
\mubold\;\cdot\;\frac{\partial(\efbold\times\gbold)}{\partial
\omega}\;=0\;\; .
\end{equation}

To continue, we note that in the two-body setting the
ratio $\;{\bf\vec J}/\mu\;$, is known to be equal to
$\;\sqrt{a(1\,-\,e^2)}\;{\bf\hat w}\;$ where $\;\bf\hat w\;$ is a
unit vector perpendicular to the unperturbed orbit's plane.
Moreover, in planet-associated noninertial axes
$\,(x,\,y,\,z)\,$ with corresponding unit vectors
$\,({\bf{\hat{x}}},\,{\bf{\hat{y}}},\,{\bf{\hat{z}}})\,$, the
normal to the orbit is expressed by
 \begin{equation}
 {\bf\hat{w}}\;=\;{\bf\hat{x}}\;\sin \inc\;\sin \Omega\;-
 \;{\bf\hat{y}}\;\sin
 \inc\;\cos \Omega\;+\;{\bf\hat{z}}\;\cos \inc\;\; .
 \label{normal}
 \end{equation}
Hence,
 \begin{equation}
 \mubold\;\cdot\;\frac{\partial\,\left(\efbold\times\gbold\right)}{\partial
 a}\;=\;\mubold\;\cdot\;{\bf\hat w}\;\frac{\partial\,\left(
 \sqrt{a(1\,-\,e^2)}\;\right)}{\partial
 a}\;=\;\frac{1}{2}\;\sqrt{\frac{1\,-\,e^2}{a}}\;\,\mu_\perp\;\;,
 \label{derivative_a}
 \end{equation}
 and
 \begin{equation}
 \mubold\;\cdot\;\frac{\partial\,\left(\efbold\times\gbold\right)}{\partial
 e}\;=\;\mubold\;\cdot\;{\bf\hat w}\;\frac{\partial\,\left(
 \sqrt{a(1\,-\,e^2)}\;\right)}{\partial
 e}\;=\;-\;\sqrt{\frac{a\;e^2}{1\,-\,e^2}}\;\,\mu_\perp\;\;,
 \label{derivative_e}
 \end{equation}
where $\mu_\perp\;=\; \mu_x\;\sin \inc\;\sin \Omega  \;-\;
 \mu_y\;\sin \inc\;\cos \Omega  \;+\;
 \mu_z\;\cos \inc$ is the orthogonal-to-orbit component of the
precession rate. The remaining two derivatives look:
 \begin{eqnarray}
 \nonumber
 \mubold\;\cdot\;\frac{\partial\,\left(\efbold\times\gbold\right)}{\partial
 \Omega}\;=\;\sqrt{a\,\left(1\,-\,e^2
 \right)\,}\;\;\mubold\;\cdot\;\frac{\partial \bf\hat w}{\partial
 \Omega}\; =~~~ \\
 \label{derivative_Omega}\\
 \nonumber
 \sqrt{a\,\left(1\,-\,e^2  \right)\,}\;\;\left\{
 \mu_x\;\sin \inc\;\cos \Omega  \;+\;
 \mu_y\;\sin \inc\;\sin \Omega
 \right\}
 \end{eqnarray}
 and
 \begin{eqnarray}
 \nonumber
 \mubold\;\cdot\;\frac{\partial\,\left(\efbold\times\gbold\right)}{
 \partial \inc}\;=\;\sqrt{a\,\left(1\,-\,e^2
 \right)\,}\;\;\mubold\;\cdot\;\frac{\partial \bf\hat w}{\partial
 \inc}\; =~~~~~~~~~~~~~~~~~~~\\
 \label{derivative_i}\\
 \nonumber
 \sqrt{a\,\left(1\,-\,e^2  \right)\,}\;\;\left\{
 \mu_x\;\cos \inc\;\sin \Omega  \;-\;
 \mu_y\;\cos \inc\;\cos \Omega  \;-\;
 \mu_z\;\sin \inc
  \right\}
 \end{eqnarray}

As for the derivatives of $\;{\bf\vec a}\,\cdot\,{\efbold}\;$, they
may be calculated directly from the expression for $\;\efbold
(\Omega,\,\omega,\,\inc,\,a,\,e,\,M_o\,;\;t) \;$ presented
above. However, the resulting expressions are cumbersome so we do not
present them here.

\pagebreak

          \end{document}